\documentclass[a4paper, prl, twocolumn,superscriptaddress,showpacs]{revtex4}

\usepackage{amssymb}
\usepackage{amsmath}
\usepackage{epsfig}
\usepackage{color}
\usepackage{graphics, graphicx}
\usepackage{bbold}
\usepackage{psfrag}
\usepackage{mathcomp}
\usepackage{subfigure}
\usepackage{verbatim}
\usepackage{color}
\usepackage[colorlinks,citecolor=blue]{hyperref}

\begin{document}

\date{\today}
\title{Topological superradiance in a degenerate Fermi gas}
\author{Jian-Song Pan}
\affiliation{Key Laboratory of Quantum Information, University of Science and Technology of China, CAS, Hefei, Anhui, 230026, China}
\affiliation{Synergetic Innovation Center of Quantum Information and Quantum Physics, University of Science and Technology of China, Hefei, Anhui 230026, China}
\author{Xiong-Jun Liu}
\email{xiongjunliu@pku.edu.cn}
\affiliation{International Center for Quantum Materials, School of Physics, Peking University, Beijing 100871, China}
\affiliation{Collaborative Innovation Center of Quantum Matter, Beijing 100871, China}
\author{Wei Zhang}
\email{wzhangl@ruc.edu.cn}
\affiliation{Department of Physics, Renmin University of China, Beijing 100872, China}
\affiliation{Beijing Key Laboratory of Opto-electronic Functional Materials and Micro-nano Devices,
Renmin University of China, Beijing 100872, China}
\author{Wei Yi}
\email{wyiz@ustc.edu.cn}
\affiliation{Key Laboratory of Quantum Information, University of Science and Technology of China, CAS, Hefei, Anhui, 230026, China}
\affiliation{Synergetic Innovation Center of Quantum Information and Quantum Physics, University of Science and Technology of China, Hefei, Anhui 230026, China}
\author{Guang-Can Guo}
\affiliation{Key Laboratory of Quantum Information, University of Science and Technology of China, CAS, Hefei, Anhui, 230026, China}
\affiliation{Synergetic Innovation Center of Quantum Information and Quantum Physics, University of Science and Technology of China, Hefei, Anhui 230026, China}

\begin{abstract}
We predict the existence of a topological superradiant state in a two-component degenerate Fermi gas in a cavity. The superradiant light generation in the transversely driven cavity mode induces a cavity-assisted spin-orbit coupling and opens a bulk gap at half filling. This mechanism can simultaneously drive a topological phase transition in the system, yielding a topological superradiant state. We map out the steady-state phase diagram in the presence of an effective Zeeman field, and identify a critical tetracritical point beyond which the topological and the conventional superraidiant phase boundaries separate. The topological phase transition can be detected from its signatures in either the momentum distribution of the atoms or the variation of the cavity photon occupation.
\end{abstract}

\maketitle

\emph{Introduction}.--
For an ultracold atomic gas inside an optical cavity, the interplay between the atomic motion and the light field can often give rise to rich dynamical processes and novel many-body phases~\cite{cavbecexp1,cavatomrev,drivendiss2,spinglass1,spinglass2,spinglass3}.
An important experimental achievement in these systems is the recent observation of the Dicke superradiance in a Bose-Einstein condensate (BEC) coupled to cavity light fields, where the backaction of the cavity photons induces an open-system version of the supersolid state in the BEC~\cite{cavbecexp1,dicketheory1,dicketheory2,dicketheory3}. For a degenerate spinless Fermi gas in a cavity, it has been shown theoretically that the atomic scattering of cavity photons in the presence of a Fermi surface can lead to the enhancement of superradiance due to the nesting effect~\cite{cavfermion1,cavfermion2,cavfermion3}. Furthermore, at the superradiant (SR) phase transition, a bulk gap opens at the Fermi surface as a result of the atom-photon scattering~\cite{cavfermion2}.

\begin{figure}[tbp]
\includegraphics[width=8cm]{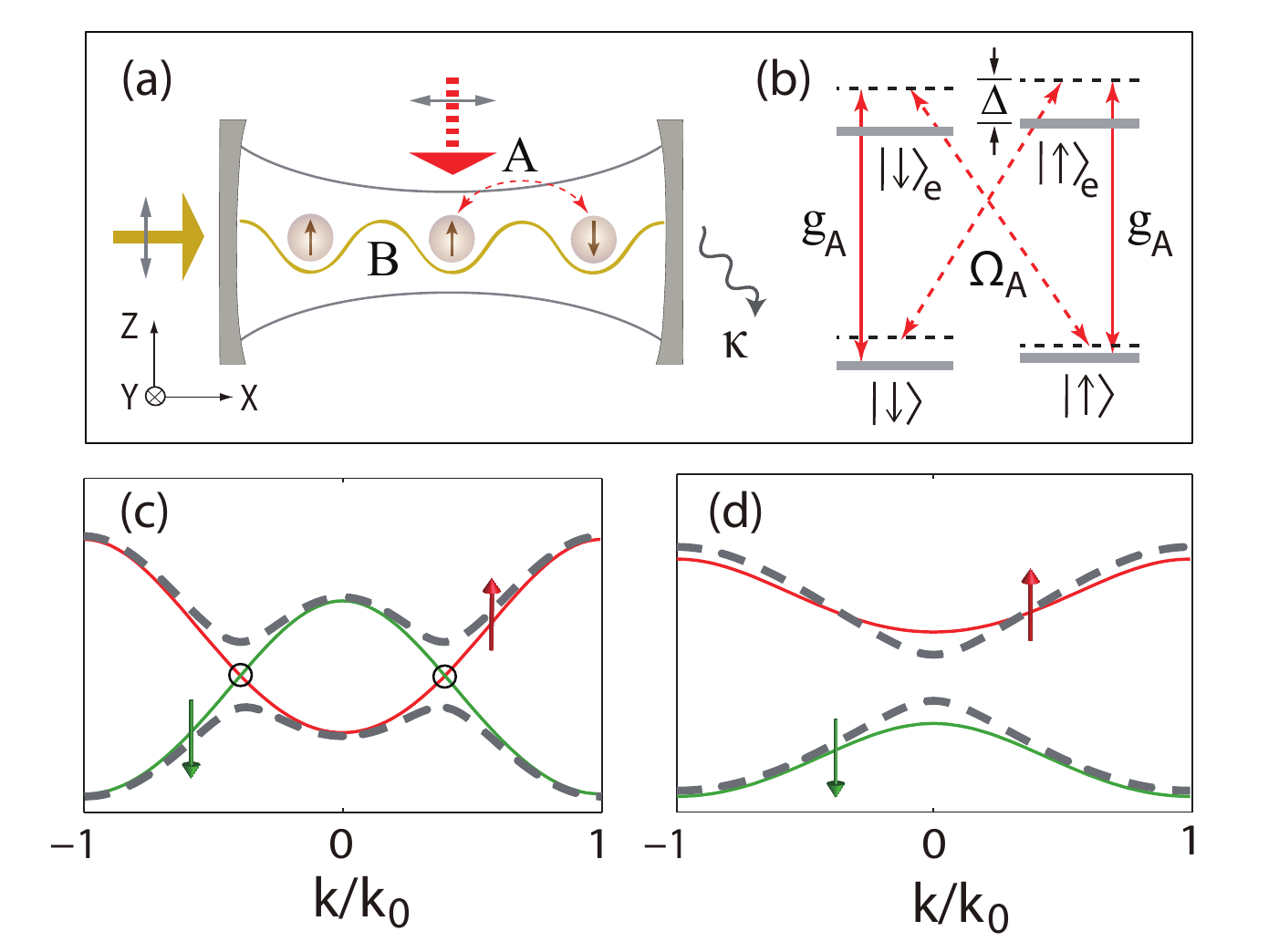}
\caption{(Color online) (a) A quasi-one-dimensional Fermi gas is coupled to a two-mode optical cavity, which is under both transverse (along $-\hat{z}$) and longitudinal (along $\hat{x}$) pumping. (b) Atomic level scheme. (c) Superradiance-induced gap opening, at positions marked by circles, for $m_z<m_c$ at half filling. (d) The persistence of the bulk gap across the SR transition at $m_z>m_c$. In (c)(d), the solid (dashed) curves are the lowest-band dispersion spectra in the first Brillouine zone before (after) the SR transition. The dispersion spectrum of the spin-down species (solid green) is shifted by half the Brillouine zone relative to that of the spin-up species (solid red) under the cavity-assisted SOC.}
\label{fig:schematic}
\end{figure}

Here, we show that for a spinful degenerate Fermi gas in a cavity, a superradiance-induced gap opening can lead to the stabilization of an exotic topological superradiant (TSR) state, where the SR light generation in the cavity is accompanied by the appearance of topologically nontrivial properties of the Fermi gas~\cite{kanereview,zhangscreview}. Characterized simultaneously by a local order parameter and a global topological invariant, this novel steady-state topological phase is intimately connected to the cavity-assisted spin-orbit coupling (SOC) inherent in the system. While the synthetic SOC in ultracold atomic systems has been extensively studied following its experimental implementation~\cite{gauge2exp,fermisocexp1,fermisocexp2,socreview1,socreview2,socreview3,socreview4,socreview5}, the cavity-assisted SOC emerges as a promising setting with rich physical implications~\cite{cavgauge1,cavgauge2}.

We focus on the superradiance of a quasi-one-dimensional, noninteracting Fermi gas with cavity-assisted Raman processes. As illustrated in Fig.~\ref{fig:schematic}(a), the two relevant cavity modes are driven, respectively, by a transverse and a longitudinal pumping laser with linear polarizations. Since the detuning between the two cavity modes are typically much larger than the energy scales of the relevant system dynamics~\cite{twomodcavexp}, we may treat them independently. The transversely driven cavity mode (mode A) and the transverse pumping laser couple two ground hyperfine states (labeled by $| \uparrow\rangle$ and $|\downarrow\rangle$) in two separate Raman processes via different electronically excited states $|\uparrow\rangle_e$ and $|\downarrow\rangle_e$ (Fig. \ref{fig:schematic}\,b). The magnetic quantum numbers of the atomic states satisfy $m_{|\sigma=\uparrow,\downarrow \rangle}=m_{|\sigma\rangle_e}$ and $m_{|\uparrow\rangle}=m_{|\downarrow\rangle}+1$. A bias magnetic field applied along the quantization axis $\hat{z}$ can provide an effective Zeeman field $m_z$ between $|\uparrow\rangle$ and $|\downarrow\rangle$. The frequency $\omega_c$ of mode A is close to that of the pumping laser $\omega_A$, both of which are blue detuned from the excited states with the single-photon detuning $\Delta\gg \Omega_A, g_A$~\cite{footnote1}. Here $\Omega_A$ is the Rabi frequency of the transverse pumping laser and $g_A$ is the single-photon Rabi frequency of mode A. Adiabatically eliminating the excited states, we get the effective Rabi frequency of the cavity-assisted Raman processes: $\eta=\Omega_Ag_A/\Delta$. The atoms are also subject to a background one-dimensional lattice potential along $\hat{x}$ due to the longitudinally pumped cavity mode (mode B). When $m_z$ is greater than a critical value $m_c$, the band gap opened by this background lattice leads to a ferromagnetic insulator (I) state at half filling prior to the SR transition. We will see that the presence of this I phase results in a rich phase diagram with an interesting tetracritical point and a novel topological phase boundary, which facilitates the observation of the TSR state.

We study the SR transition in mode A when the transverse pumping and the effective Zeeman field are tuned. Typically, when the strength of the transverse pumping increases beyond a critical value, the system becomes superradiant, with a macroscopic photon occupation in mode A and the onset of a finite order parameter. We show that the effects of the backaction of cavity photons on the Fermi gas depend crucially on the initial state of the fermions, and are particularly important when the background lattice is half filled.

For $m_z<m_c$, the Fermi gas is in a gapless metallic (M) state at small pumping. The SR transition induces a gap at finite momenta out of the M phase (Fig.~\ref{fig:schematic}\,c). This is because the spatially varying cavity field of mode A has twice the period of that of the background potential, so that the Raman processes effectively play the role of a cavity-assisted SOC which couples fermions with different hyperfine spins and with a momentum difference that spans half of the Brillouine zone. Importantly, we show that the SR light generation in mode A and the bulk gap opening are accompanied by the change of topological properties of the Fermi gas. The system is thus in an exotic TSR phase, whose topological nature can be confirmed by edge-state and winding-number calculations.

For $m_z>m_c$, the ferromagnetic bulk gap persists as the system crosses the SR transition (Fig.~\ref{fig:schematic}\,d). The Fermi gas thus changes from an I state to a topologically trivial SR state across the SR transition. However, we find that by further increasing the pumping strength, the bulk gap closes and opens up again at zero momentum ($k=0$), as the system crosses a topological phase boundary to become TSR. By mapping out the steady-state phase diagram, we show that the TSR state here can be adiabatically connected to the TSR state at $m_z<m_c$, while a tetracritical point exists where different phase boundaries merge. The topological phase transition between TSR and SR can be detected either from the spin-resolved momentum distribution of the Fermi gas or from the variation of the cavity-photon occupation.

\emph{Model}.--
Adopting the mean-field approximation on the cavity fields, we look for the steady-state solution of the driven-dissipative system, where the mean fields of the cavity modes are stationary. Integrating out the tightly confined transverse degrees of freedom, we have the effective one-dimensional Hamiltonian~\cite{supple}:
\begin{align}
\hat{H}=&\sum_{\sigma}\int dx \hat{\psi}^{\dag}_{\sigma}\Big[
\frac{p_x^2}{2m}+(V_0+\xi_A|\alpha|^2)\cos^2(k_0x)
+\xi_{\sigma}m_z\Big]
\hat{\psi}_{\sigma}\nonumber\\
&+\eta_A(\alpha^{\ast}+\alpha)\left[\int dx \hat{\psi}^{\dag}_{\uparrow}\cos(k_0x)\hat{\psi}_{\downarrow}+\text{H.C.}\right],
\label{eqn:effH}
\end{align}
where $\hat{\psi}_{\sigma}$ ($\sigma=\uparrow,\downarrow$) are the effective one-dimensional fermionic field operators for different hyperfine states, $\xi_{\sigma}=\pm 1$, $m$ is the atomic mass, and $\text{H.C.}$ represents Hermitian conjugate. The atoms are subject to an effective lattice potential $(V_0+\xi_A|\alpha|^2)\cos^2(k_0x)$, where $\xi_{A}=g_A^2/\Delta$, and $V_0\cos^2(k_0x)$ is the background potential generated by mode B. The wave vectors of the cavity modes are approximately the same, which we denote as $k_0$. The effective pumping strength $\eta_A=s\eta$, with the constant $s$ coming from the transverse integrals. The cavity mean field $\alpha$ for mode A can be determined by the stationary condition $\partial\alpha/\partial t=0$, which is proportional to the order parameter $\Theta=\int dx\theta(x)=\int dx\left(\langle\hat{\psi}^{\dag}_{\downarrow}\hat{\psi}_{\uparrow}\rangle+\text{H.C.}\right)\cos(k_0x)$~\cite{supple}. We will use the recoil energy $E_r=\hbar^2k_0^2/2m$ as the unit of energy in the following discussions.

\begin{figure}[tbp]
\includegraphics[width=8.5cm]{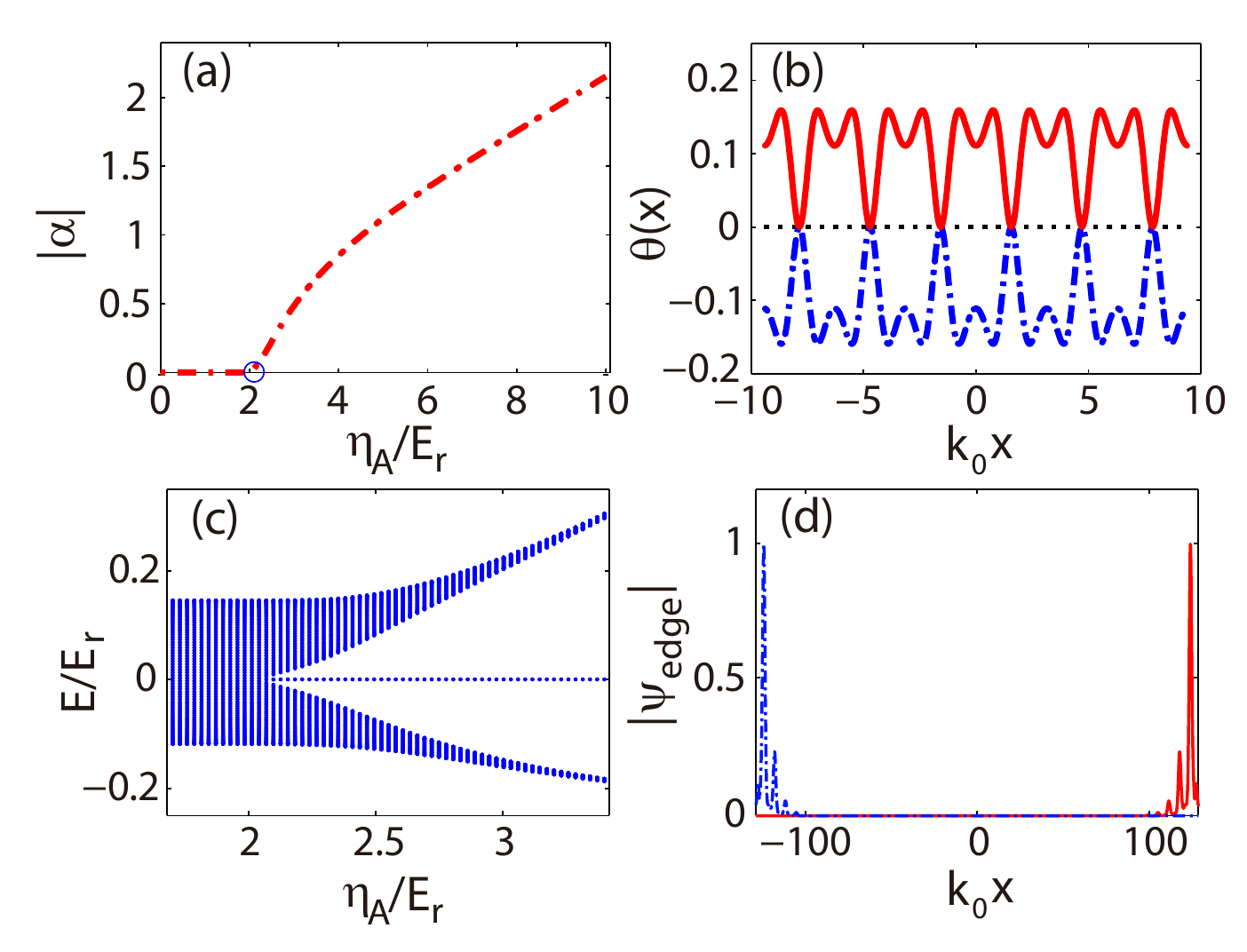}
\caption{(Color online) TSR in a quasi-one-dimensional lattice with open boundary conditions. (a) The cavity field $|\alpha|$ across the TSR phase transition with increasing $\eta_A$. (b) $\theta(x)$ (see main text) before (dotted curve, with $\eta_A= 1 E_r$) and after (solid and dash-dotted curves, with $\eta_A=3 E_r$) the TSR phase transition on the central $6$ sites. The critical point is at $\eta_A^c\sim 2.05E_r$. The solid (dash-dotted) curve corresponds to the TSR phase whose cavity field acquires a positive (negative) real part due to the spontaneous symmetry breaking. (c) Edge state emerges in the superradiance-induced bulk gap as the system crosses the phase boundary. (d) Wave functions for the edge states in (c) at $\eta_A=3E_r$. In all cases, we consider a half-filled lattice of $80$ sites, with the parameters: $k_BT=E_r/200$, $m_z=0$, $V_0=5E_r$, $\kappa= 100E_r$, $\Delta_A=-10E_r$, $\xi_A= 5 E_r$. For $^6$Li atoms, these can be satisfied by choosing: $\kappa\sim 7.4$MHz, $g_A\sim 27.1$MHz, $|\Delta_A|\sim 0.74$MHz, $\Delta\sim 2$GHz, $T\sim 17.7$nK.}
\label{fig:topochara}
\end{figure}

\emph{Topological superradiance}.--
An outstanding feature of the system under Hamiltonian (\ref{eqn:effH}) is the existence of an SR transition when the effective transverse pumping strength $\eta_A$ increases. The critical pumping strength for the SR transition can be derived by evaluating the free energy using second-order perturbation~\cite{cavfermion3}. Integrating out the fermion fields, and requiring a vanishing coefficient for the second-order expansion in $\Theta$ of the free energy, the critical pumping strength is given by:
\begin{align}
\eta^c_A=\frac{1}{2}\sqrt{\frac{\tilde{\Delta}_A^2+\kappa^2}{-\tilde{\Delta}_A f}},\label{eqn:omegac}
\end{align}
where $\tilde{\Delta}_A=\Delta_A-\xi_A\sum_{j,\sigma}\int dx |\varphi_{j\sigma}|^2\cos^2(k_0x) n_{F}(\epsilon_j)$,  the cavity detuning $\Delta_A=\omega_A-\omega_c$, $f=\frac{1}{2}{\sum_{j,j'}}|M_{jj'}|^2 [n_F(\epsilon_{j'})-n_F(\epsilon_{j})]/(\epsilon_{j}-\epsilon_{j'})$, and $\kappa$ is the cavity decay rate.
Here, $M_{jj'}=\sum_{\sigma\neq\sigma'}\int dx\varphi^{\ast}_{j\sigma}\cos(k_0x)\varphi_{j'\sigma'}$, the Fermi-Dirac distribution $n_F(x)=1/(e^{(x-\mu)/k_BT}+1)$ with chemical potential $\mu$ and temperature $T$, and $k_B$ is the Boltzmann constant. $\{\varphi_{j\uparrow},\varphi_{j\downarrow}\}^T$ is the eigen state of the Hamiltonian $p_x^2/2m+V_0\cos^2k_0x+ m_z\sigma_z$ with eigen energy $\epsilon_{j}$, and $\sigma_z$ is the Pauli matrix.

Prior to the SR transition, $\alpha=0$ and the atoms experience the lattice potential $V_0\cos^2(k_0x)$. In the SR regime, the macroscopic photon occupation of mode A modifies the background lattice potential, and introduces a spatially varying cavity-assisted SOC term in the effective Hamiltonian (\ref{eqn:effH}) with twice the period as that of the background potential. When the cavity mode A is weak, we can neglect the small spin-dependent interband coupling. Then, taking the gauge transformation $\left\{\hat{\psi}_{\uparrow}\rightarrow\hat{\psi}_{\uparrow}, \hat{\psi}_{\downarrow}\rightarrow-ie^{ik_0x}\hat{\psi}_{\downarrow}\right\}$ in the single-band tight-binding limit, we can map the Hamiltonian (\ref{eqn:effH}) onto the one underlying a chiral topological insulator, which is topologically nontrivial below a critical Zeeman field at half filling~\cite{xiongjunprl13}. When $\eta_A$ increases, the interband couplings become appreciable, and the single-band tight-binding approximation no longer applies. Nevertheless, we will show that the topological properties persist even in the deep SR regime.

We first diagonalize the effective Hamiltonian (\ref{eqn:effH}) at $m_z=0$ for a finite-size lattice under open boundary conditions, while solving the cavity field self-consistently. The resulting cavity field and energy spectrum of the bulk states are shown in Fig.~\ref{fig:topochara}\,a and Fig.~\ref{fig:topochara}\,c, respectively. Apparently, as soon as the pumping exceeds a critical value, a pair of zero modes with localized wave functions (Fig.~\ref{fig:topochara}\,d) emerge in the superradiance-induced bulk gap, and a finite order parameter $\Theta$ appears simultaneously (Fig. \ref{fig:topochara}\,b). The SR transition here is then topological as well, and the system is in the TSR phase beyond this TSR phase transition.

\begin{figure}[tbp]
\includegraphics[width=8.5cm]{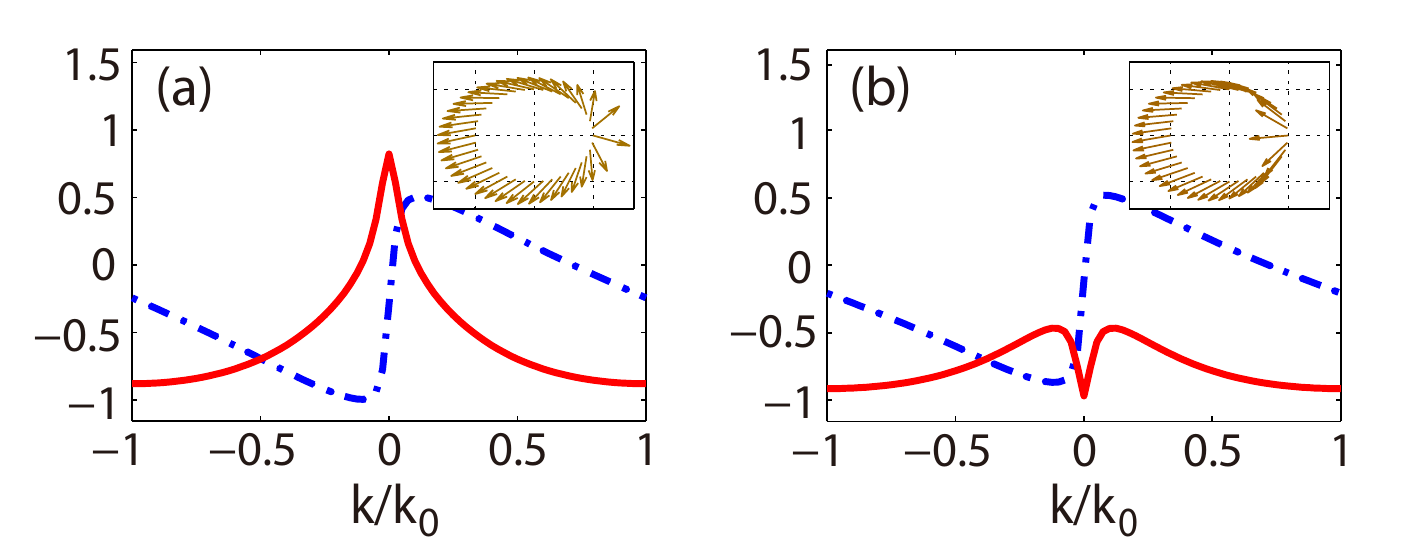}
\caption{(Color online) Typical momentum distribution of $\langle\sigma_z\rangle$ (solid) and $\langle\sigma_y\rangle$ (dashed) in the (a) TSR and (b) SR phases at $\eta_A=3E_r$. Inset: Corresponding spin textures $\langle\vec{\sigma}\rangle$ in the first Brillouin zone (shown in a loop). The effective Zeeman fields are (a) $m_z=0.171E_r$ and (b) $m_z=0.181E_r$, where the topological phase transition occurs at $m_z=0.176E_r$ for $\eta_A= 3 E_r$. Other parameters are the same as those in Fig.~\ref{fig:topochara}.}\label{fig:winding}
\end{figure}

\emph{Winding number and steady-state phase diagram}.--
To characterize the stability of the TSR phase, we examine the steady-state phase diagram on the $\eta_A$--$m_z$ plane. While the SR phase boundaries are calculated from Eq.~(\ref{eqn:omegac}), the topological phase boundaries can be determined by calculating the winding number, which serves as the bulk topological invariant.

The bulk topological invariant can be defined by a mapping from the first Brillouine zone to the spin space, and can be read out from the momentum-space spin texture $\langle\vec{\sigma}\rangle_k=\langle\sigma_y\rangle_k \vec{e}_y+\langle\sigma_z\rangle_k \vec{e}_z$, where $\sigma_y$, $\sigma_z$ are Pauli matrices. The expectation value is taken with respect to the lowest-band Bloch state in momentum space, which is obtained by solving Hamiltonian (\ref{eqn:effH}) self-consistently. Note that the spin of a Bloch state lies in the $y$--$z$ plane, with the unit vector along the spin orientation being an element of a closed circle $S^1$. The winding number of the insulating state can be calculated by counting the number of times $S^1$ is covered by the mapping. The momentum-space spin textures of a finite-size system under different Zeeman fields are shown in Fig.~\ref{fig:winding}. We find that, depending on $\eta_A$ and $m_z$, the winding number can be either $1$ (inset of Fig.~\ref{fig:winding}\,a) or $0$ (inset of Fig.~\ref{fig:winding}\,b) in the superradiant region, corresponding to the TSR state or a trivial SR insulator. Across the TSR--SR phase boundary, the bulk gap closes and opens up again at $k=0$, while the spin texture at $k=0$ undergoes an abrupt change (Fig.~\ref{fig:winding}\,a,b), indicating a topological phase transition at the boundary.

\begin{figure}
\includegraphics[width=8.0cm]{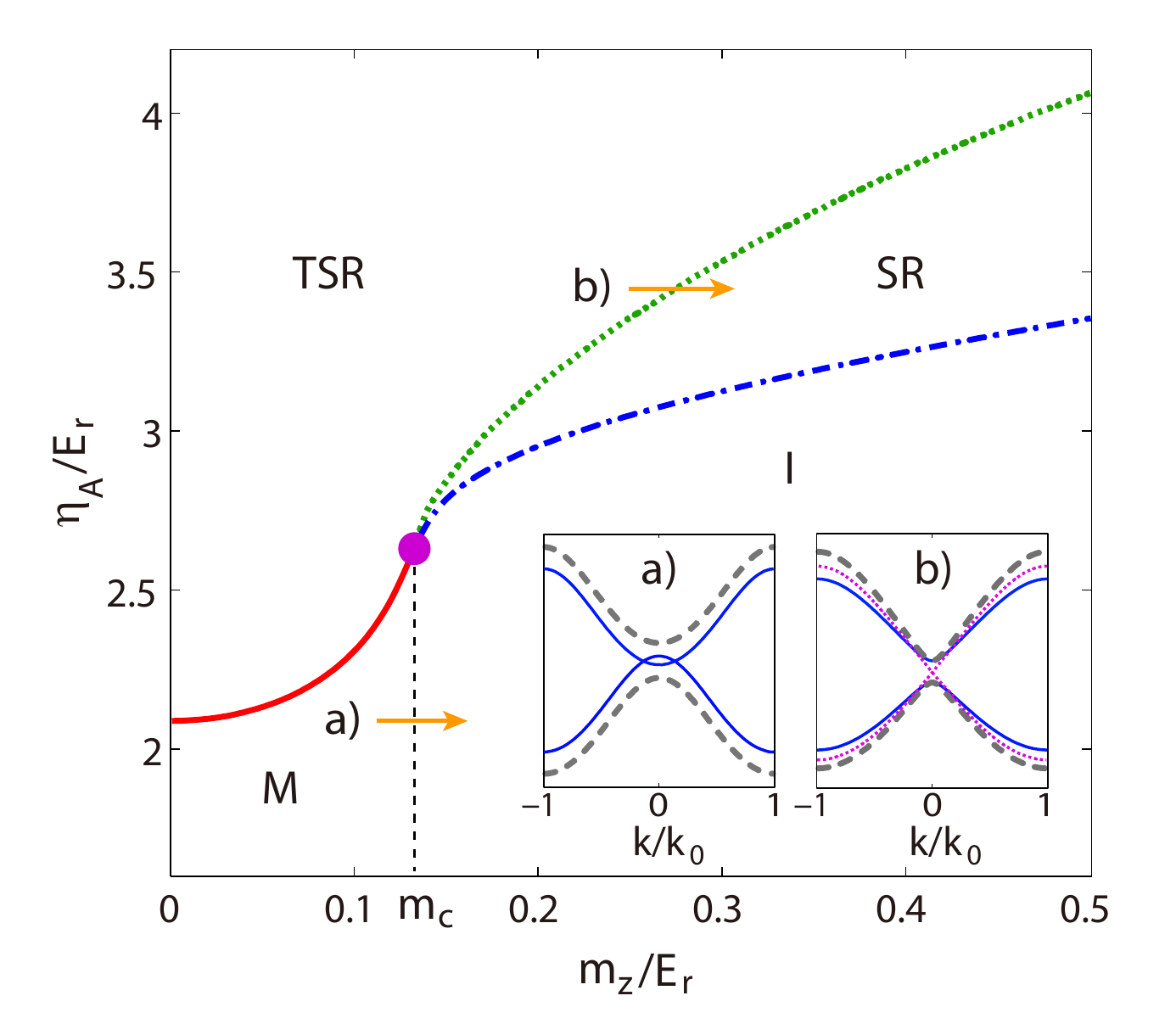}
\caption{(Color online) Steady-state phase diagram at a finite temperature $k_BT =E_r/200$. The solid curve is the TSR phase boundary, and the dotted curve is the topological phase boundary between the TSR and the trivial SR states. The thin dashed curve at $m_c\sim0.132E_r$ is the boundary between the M and the I states, and the dash-dotted curve is the conventional SR phase boundary. The various boundaries merge at a tetracritical point (dot) at $\eta_A\sim 2.614 E_r$, $m_c\sim 0.132E_r$. Other parameters are the same as those used for Fig.~\ref{fig:topochara}. Inset: change of bulk gap before (solid), after (dashed) and right at (dotted) the phase boundaries labeled by arrows.}
\label{fig:phasediag}
\end{figure}

We then map out the steady-state phase diagram in Fig.~\ref{fig:phasediag}. An important observation here is that the TSR--SR boundary always occurs at $m_z>m_c$. We emphasize that this topological phase transition is due to the deformation of the ferromagnetic band structure driven by Raman-assisted interband couplings, which are dominant when $\eta_A$ is large. More specifically, as $\eta_A$ increases from the I state at $m_z>m_c$, the system first crosses an I--SR boundary to become a trivial SR state. Further increasing $\eta_A$ enhances the interband couplings, which close the bulk gap at the TSR-SR boundary by pushing downward the lowest subbands~\cite{supple}. Interestingly, due to the band inversion at the gap closing point, the interband couplings would then enlarge the bulk gap as $\eta_A$ increases toward the TSR regime. As a remarkable consequence, a tetracritical point appears on the phase diagram, across which different types of phase transitions can be observed along different lines in the $\eta_A$--$m_z$ plane~\cite{footnote2}.

\begin{figure}
\includegraphics[width=8.5cm]{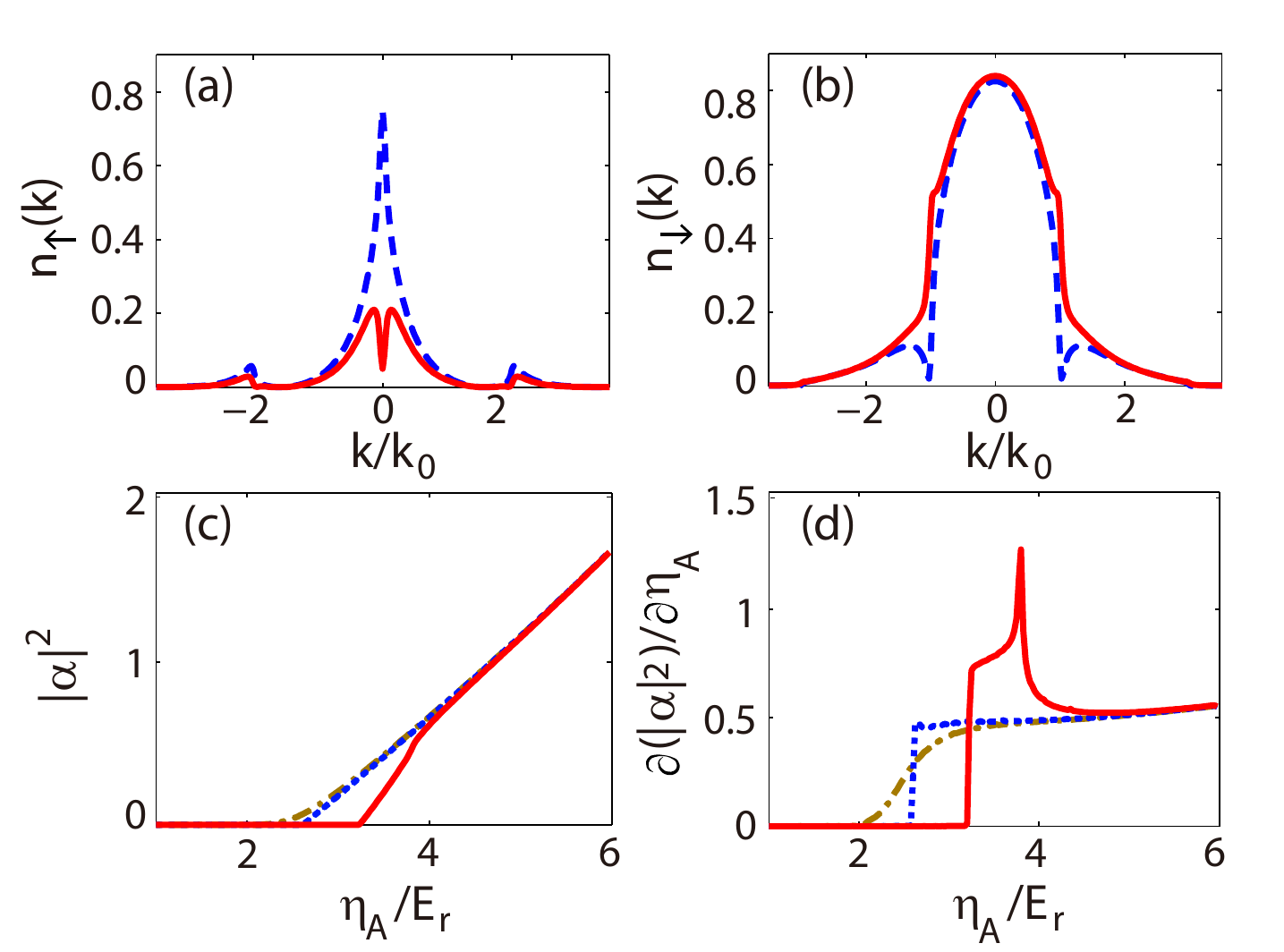}
\caption{(Color online) (a)(b) Momentum distribution of the spin-up (a) and the spin-down (b) atoms in the TSR state (dashed) and the trivial SR state (solid), respectively. The parameters of the dashed (solid) curves are the same as those in Fig.~\ref{fig:winding}\,a (Fig.~\ref{fig:winding}\,b). (c)(d) Variation of the cavity-photon number across the TSR-SR topological phase boundary on the phase diagram in Fig.~\ref{fig:phasediag}, with $m_z=0.03E_r$ (dash-dotted), $m_z=0.132E_r$ (dotted), $m_z=0.4E_r$ (solid).}
\label{fig:detection}
\end{figure}

\emph{Detection}.--
The topological phase boundary is characterized by abrupt changes in winding number and the closing of the bulk gap, both of which leave detectable signatures in our system. Firstly, the abrupt change in $\langle\sigma_z\rangle_k$ at the gap closing point in momentum space across the TSR--SR phase boundary is uniquely associated with the change of winding number (see Fig.~\ref{fig:winding}).
This leads to an abrupt change in the momentum-space density distribution of the spin-up fermions at $k=0$ and that of the spin-down atoms at $|k|=k_0$ across the topological phase boundary (see Fig.~\ref{fig:detection}\,a,b). This abrupt change can in principle be detected using the spin-selective time-of-flight imaging technique. Secondly, at the gap closing point, a spin-up fermion with $k=0$ can be scattered by a cavity photon into a spin-down state at $|k|=k_0$ with the same energy. This nesting condition at the gap closing point leads to a peak in the derivative of the cavity photon number at the topological phase boundary (see Fig.~\ref{fig:detection}\,c,d). Therefore, by monitoring the photons leaking out of the cavity, one may detect both the SR phase transition and the TSR--SR topological phase transition. We have checked that both signals are robust against small Fermi-energy deviations at higher temperatures~\cite{supple}.

\emph{Final remarks}.--
Experimentally, quasi-one-dimensional Fermi gases can be prepared by imposing a two-dimensional lattice potential with tight transverse confinement, leading to an array of quasi-one-dimensional atomic gases with large number of atoms. With more atoms in the cavity, the TSR phase and the corresponding phase transition can emerge at more favorable parameters such as weaker atom-cavity couplings, larger single-photon detunings, and larger cavity decay rates~\cite{supple}. Finally, we note that our scheme can be straightforwardly extended to quasi-two-dimensional Fermi gases in a cavity, where a TSR phase should also exist, as the single-band tight-binding Hamiltonian just beyond the SR transition therein can be mapped onto that of a quantum anomalous Hall system~\cite{xiongjunprl14}.

\emph{Acknowledgments}.-- We thank Yu Chen, Dong-Sheng Ding, Su Wang, Zhen-Biao Yang, and Hui Zhai for helpful discussions.
This work is supported by NFRP (2011CB921200, 2011CBA00200), NKBRP (2013CB922000),
NNSF (60921091), NSFC (11105134, 11274009, 11374283), the Fundamental Research Funds
for the Central Universities (WK2470000006), and the Research Funds of
Renmin University of China (10XNL016).

\newpage

\begin{widetext}
\appendix
\section{Supplemental Materials}

\subsection{\textbf{A. Effective Hamiltonian}}\label{Effective Hamiltonian}

In this section, we provide more details on the derivation of the effective Hamiltonian Eq.~(1) in the main text. We consider the typical case where the two relevant cavity modes, mode A and mode B, have a frequency difference of hundreds of MHz~\cite{subtwomodcavexp}. Mode B is then effectively decoupled from the relevant dynamics of the system due to the large detuning. The effect of mode B is a stationary background lattice potential $V_0\cos^2(k_0 x)$, where $k_0$ is the wave vector of mode B, which only differs from that of mode A by a negligibly small amount. In the following, we will use $k_0$ for the wave vector of mode A as well.

For a quasi-one-dimensional cavity Fermi gas illustrated in Fig.~1 in the main text, the internal dynamics of a single particle under the cavity mode A and the transverse pumping field can be described by the Hamiltonian
\begin{equation*}\label{equation S1}
\mathcal{H}=\mathcal{H}_{0}+\mathcal{H}_{I},
\tag{S1}
 \end{equation*}
 where
\begin{equation*}\label{equation S2}
\mathcal{H}_{0}=\hbar\omega_{c}\hat{a}^{\dagger}\hat{a}+\sum_{\sigma=\uparrow,\downarrow}\hbar\omega_{\sigma e}|\sigma\rangle_e{}_e\langle \sigma|+\sum_{\sigma=\uparrow,\downarrow}\hbar\omega_{\sigma}|\sigma\rangle_g{}_g\langle\sigma|,
\tag{S2}
\end{equation*}
\begin{equation*}\label{equation S3}
\mathcal{H}_{I}=-\left(\Omega_{1\uparrow}\sigma_{1\uparrow}^{+}+\Omega_{2\downarrow}\sigma_{2\downarrow}^{+}\right)e^{-i\left(\omega_{A}t+k_{0}z\right)}-\left(g_{1\downarrow}\sigma_{1\downarrow}^{+}\hat{a}+g_{2\uparrow}\sigma_{2\uparrow}^{+}\hat{a}\right)\cos\left(k_{0}x\right)+\text{H.C.}
\tag{S3}
\end{equation*}
Here $\omega_{\sigma e}$ ($\omega_{\sigma}$) denotes the eigen frequency of the exited states $|\sigma\rangle_e$ (the ground states $|\sigma\rangle_g$). The ground states are typically split by an effective Zeeman field, i.e., $\omega_{\uparrow}=-\omega_{\downarrow}=m_z$. We assume the Zeeman field to be along the $\hat{z}$ axis, which defines the quantization axis. The resonant frequency of mode A is denoted as $\omega_{c}$. The linearly polarized transverse pumping field propagates in the $-\hat{z}$ direction and has a frequency of $\omega_{A}$. The raising operators $\sigma_{1\sigma}^{+}=|\downarrow\rangle_e{}_g\langle \sigma|$, $\sigma_{2\sigma}^{+}=|\uparrow\rangle_e{}_g\langle \sigma|$, and $\text{H.C.}$ stands for Hermitian conjugate. The two Raman channels are respectively driven by the two circularly polarized components of the transverse pumping field with the effective Rabi frequency $\Omega_{j\sigma}$ ($j=1,2$). $g_{j\sigma}$ ($j=1,2$) denotes the single-photon Rabi frequency of the cavity-atom coupling. For simplicity, we assume  $g_{1\downarrow}=g_{2\uparrow}=g_{A}$ and $\Omega_{1\uparrow}=\Omega_{2\downarrow}=\Omega_{A}$, which can always be satisfied by adjusting the system parameters for the commonly used fermion species such as $^6$Li and $^{40}$K.

Introducing the time-dependent rotation, $U\left(t\right)=\exp\left[i\left(\sum_{\sigma=\uparrow,\downarrow}|\sigma\rangle_e{}_e\langle \sigma|+\hat{a}^{\dagger}\hat{a}\right)\omega_{A}t\right]$, and adiabatically eliminating the two exited states, we have
\begin{equation*}\label{equation S4}
\mathcal{H}_{eff}=-\left[\Delta_{A}-\xi_{A}\cos^{2}\left(k_{0}x\right)\right]\hat{a}^{\dagger}\hat{a}+\hbar m_{z}\sigma_{z}+\eta\left(\hat{a}|\uparrow\rangle\langle\downarrow|e^{ik_{0}z}+\text{H.C.}\right)\cos\left(k_{0}z\right),
\tag{S4}
\end{equation*}
where $\Delta_{A}=\omega_{c}-\omega_{A}$ is the cavity detuning, $\xi_{A}=g_{A}^{2}/\Delta$, and the effective Rabi frequency of the cavity-assisted Raman processes $\eta=\Omega_{A}g_{A}/\Delta$. Here, the single-photon detuning $\Delta=\omega_{\uparrow e}-\omega_{A}\approx\omega_{\downarrow e}-\omega_{A}$. Note that with the adiabatic elimination of the excited states,  we have now dropped the subscript for the ground states, so that $|\sigma\rangle_g=|\sigma\rangle$.

Including the kinetic energy term, the radial trapping potential $U\left(\boldsymbol{r}\right)$ and the background lattice potential $V_{0}\cos^{2}\left(k_{0}x\right)$, we can write the effective Hamiltonian in a second-quantization form
\begin{equation*}\label{equation S5}
\begin{split}
\hat{H}=&\sum_{\sigma}\int d\boldsymbol{r}\hat{\Psi}_{\sigma}^{\dagger}\left[\frac{\boldsymbol{p}^{2}}{2m}+U\left(\boldsymbol{r}\right)+\left(V_{0}+\xi_{A}\hat{a}^{\dagger}\hat{a}\right)\cos^{2}\left(k_{0}x\right)+\xi_{\sigma}m_{z}\right]\hat{\Psi}_{\sigma}-\Delta_{A}\hat{a}^{\dagger}\hat{a}\\
&+\eta\left[\int d\boldsymbol{r}\hat{\Psi}_{\uparrow}^{\dagger}\left(\hat{a}e^{ik_{0}z}+\hat{a}^{\dagger}e^{-ik_{0}z}\right)\cos\left(k_{0}x\right)\hat{\Psi}_{\downarrow}+\text{H.C.}\right].
\end{split}
\tag{S5}
\end{equation*}
For a quasi-one-dimensional gas with tight radial confinement in the $y$--$z$ plane, we assume that only the ground state of the radial degree of freedom is occupied. The field operator can then be written as $\hat{\Psi}_{\sigma}\left(\boldsymbol{r}\right)=\sqrt{\frac{2}{\pi\rho^{2}}}\hat{\psi}_{\sigma}\left(x\right)\exp\left(-\frac{y^{2}+z^{2}}{\rho^{2}}\right)$, where $\rho$ is the characteristic width of the radial harmonic confinement. Integrating out the radial degrees of freedom, the effective one-dimensional Hamiltonian is given by
\begin{equation*}\label{equation S6}
\begin{split}
\hat{H}=&\sum_{\sigma}\int dx\hat{\psi}_{\sigma}^{\dagger}\left[\frac{p_{x}^{2}}{2m}+\left(V_{0}+\xi_{A}\hat{a}^{\dag}\hat{a}\right)\cos^{2}\left(k_{0}x\right)+\xi_{\sigma}m_{z}\right]\hat{\psi}_{\sigma}-\Delta_A\hat{a}^{\dagger}\hat{a}\\
&+\eta_{A}\left(\hat{a}+\hat{a}^{\dag}\right)\left[\int dx\hat{\psi}_{\uparrow}^{\dagger}\cos\left(k_{0}x\right)\hat{\psi}_{\downarrow}+\text{H.C.}\right],
\end{split}
\tag{S6}
\end{equation*}
where $\eta_{A}=s\eta$ with a ratio $s=e^{-k_{0}^{2}\rho^{2}/8}$, which reduces to $1$ when the system is exactly one dimensional.

We then write down the equations of motion for $\hat{a}$
\begin{equation*}\label{equation S7}
i\dot{\hat{a}}=\hat{a}\left[\xi_{A}\sum_{\sigma}\int dx\hat{\psi}_{\sigma}^{\dagger}\cos^{2}\left(k_{0}x\right)\hat{\psi}_{\sigma}-\Delta_{A}-i\kappa\right]+\eta_{A}\left[\int dx\hat{\psi}_{\downarrow}^{\dagger}\cos\left(k_{0}x\right)\hat{\psi}_{\uparrow}+\text{H.C.}\right].
\tag{S7}
\end{equation*}
Taking the mean field approximation $\langle\hat{a}\rangle=\alpha$ and imposing the stationary condition $\partial\alpha/\partial t=0$, we have the steady-state cavity field
\begin{equation*}\label{equation S8}
\alpha\approx\frac{\eta_{A}\int dx\cos\left(k_{0}x\right)\left[\langle\hat{\psi}_{\downarrow}^{\dagger}\hat{\psi}_{\uparrow}\rangle+\text{H.C.}\right]}{\Delta_{A}+i\kappa-\xi_{A}\sum_{\sigma}\int dx\langle\hat{\psi}_{\sigma}^{\dagger}\hat{\psi}_{\sigma}\rangle\cos^{2}\left(k_{0}x\right)},
\tag{S8}
\end{equation*}
where $\kappa$ is the cavity decay. Replacing the cavity field operators with their mean fields in Eq.~(\ref{equation S8}), we arrive at the effective Hamiltonian (1) in the main text.

The self-consistent calculation of cavity field can be divided into following several steps:
1. Solve Hamiltonian (Eq.~(\ref{equation S6})) by substituting an input cavity field $\alpha_0$ for the cavity operator $\hat{a}$;
2. Solve for the chemical potential from the number equation, $N=\sum_{\sigma}\int dx\langle\hat{\psi}_{\sigma}^{\dagger}\hat{\psi}_{\sigma}\rangle$;
3. Calculate the cavity field $\alpha$ with Eq.~(\ref{equation S8});
4. Stop the calculation when $\left|\alpha-\alpha_{0}\right|<\epsilon$, where $\epsilon$ is a given precision. Otherwise, update $\alpha_0$ with $\alpha$ and repeat 1-4.

\subsection{\textbf{B. Superradiant Critical Point}}\label{Critical Point of Superradiance}

The critical point of the superradiant phase transition can be determined from the second-order perturbation theory. Integrating out the fermion fields and replacing the cavity field with its steady-state value Eq.~(\ref{equation S8}), we derive the free energy to the second order of the order parameter $\Theta=\int dx\cos\left(k_{0}x\right)\left[\langle\hat{\psi}_{\downarrow}^{\dagger}\hat{\psi}_{\uparrow}\rangle+\text{H.C.}\right]$
\begin{equation*}\label{equation S9}
F_{\alpha}=-\left[\frac{\tilde{\Delta}_{A}}{\tilde{\Delta}_{A}^{2}+\kappa^{2}}+\chi_{\eta}\frac{4\tilde{\Delta}_{A}^{2}}{\left(\tilde{\Delta}_{A}^{2}+\kappa^{2}\right)^{2}}\right]\left(\eta_A\Theta\right)^{2}.
\tag{S9}
\end{equation*}
Here the effective cavity detuing $\tilde{\Delta}_{A}$ and the susceptibility $\chi_{\eta}$ are given by
\begin{equation*}\label{equation S10}
\tilde{\Delta}_{A}=\Delta_{A}-V_{0}\sum_{j}V_{jj}n_{F}\left(\epsilon_{j}\right),
\tag{S10}
\end{equation*}
\begin{equation*}\label{equation S11}
\chi_{\eta}=\eta_{A}^{2}f=-\frac{\eta_{A}^{2}}{2}\sum_{j,j^{'}}\left|M_{jj^{'}}\right|^{2}\frac{n_{F}\left(\epsilon_{j}\right)-n_{F}\left(\epsilon_{j^{'}}\right)}{\epsilon_{j}-\epsilon_{j^{'}}}
,\tag{S11}
\end{equation*}
where $\varphi_{j}\left(x\right)=\{\varphi_{j\uparrow},\varphi_{j\downarrow}\}^T$ is the eigen function of the Hamiltonian $\hat{H}_{0}=p_{x}^{2}/2m+V_{0}\cos^{2}\left(k_{0}x\right)+m_{z}\sigma_{z}$ with an eigen energy $\epsilon_j$, the Fermi-Dirac distribution $n_{F}\left(x\right)
 =1/\left[e^{\left(x-\mu\right)/k_{B}T}+1\right]$ with chemical potential $\mu$ and temperature $T$. The matrix elements $V_{jj'}$ and $M_{jj'}$ are given as
\begin{equation*}\label{equation S12}
\begin{split}
V_{jj^{'}}&=\sum_{\sigma}\int dx
\varphi_{j\sigma}^{\ast}\left(x\right)\cos^{2}\left(k_{0}x\right)\varphi_{j^{'}\sigma}\left(x\right),\\
M_{jj'}&={\displaystyle\sum_{\sigma\neq\sigma'}}\int dx\varphi^{\ast}_{j\sigma}\cos(k_0x)\varphi_{j'\sigma'}
\end{split}
\tag{S12}
\end{equation*}

The critical pumping strength for the superradiant phase transition can be derived by requiring a vanishing second-order expansion coefficient of the free energy
\begin{equation*}\label{equation S13}
\eta_{A}^{c}=\frac{1}{2}\sqrt{\frac{\tilde{\Delta}_{A}^{2}+\kappa^{2}}{-\tilde{\Delta}_{A}f}},
\tag{S13}
\end{equation*}
where $f$ is defined through Eq.~(\ref{equation S11}).

\subsection{\textbf{C. Characterizing the topological state}}\label{Topology Analysis}
\subsubsection{\textbf{1. Single-band tight-binding limit}}
Beyond the superradiant critical point, both the cavity field $\alpha$ and the cavisty-assisted Raman term $\eta_{A}\left(\alpha+\alpha^{\ast}\right)$ become finite. From the effective one-dimensional Hamiltonian Eq.~(1), the cavity-assisted Raman term induces a spin flip with spatial-dependent rate. When the system just crosses the critical point, the Raman term is much smaller than the initial lattice depth $V_{0}$ and the tight-binding approximation is applicable at low temperatures. Furthermore, the interband spin-dependent coupling, such as the on-site spin flip etc., is also small. Dropping these interband coupling terms, we can write the single-band tight-binding Hamiltonian corresponding to Eq.~(1) in the main text as
\begin{equation*}\label{equation S14}
\hat{H}_{TI}=-\sum_{\left\langle i,j\right\rangle ,\sigma}t_{s}^{1,1}(i,j)\hat{\psi}_{i\sigma}^{\dagger}\hat{\psi}_{j\sigma}+\sum_{\left\langle i,j\right\rangle }\left(t_{so}^{1,1}(i,j)\hat{\psi}_{i\uparrow}^{\dagger}\hat{\psi}_{j\downarrow}+\text{H.C.}\right)+m_{z}\sum_{i,\sigma}\xi_{\sigma}\hat{\psi}_{i\sigma}^{\dagger}\hat{\psi}_{i\sigma},
\tag{S14}
\end{equation*}
where $\hat{\psi}_{nj\sigma}$ is the atomic annihilation operator of spin $\sigma$ on site $j$, $\xi_{\uparrow,\downarrow}=\pm1$, and
\begin{equation*}\label{equation S15}
\begin{split}
&t_{s}^{n,n^{'}}(i,j)=\int dx\phi_{ni}^{\ast}\left(x\right)\left[\frac{\hat{p}_{x}^{2}}{2m}+\left(V_{0}+\xi_{A}\left|\alpha\right|^{2}\right)\cos^{2}\left(k_{0}x\right)\right]\phi_{n^{'}j}\left(x\right), \\
&t_{so}^{n,n^{'}}(i,j)=\eta_{A}\left(\alpha^{\ast}+\alpha\right)\int dx\phi_{ni}^{\ast}\left(x\right)\cos\left(k_{0}x\right)\phi_{n^{'}j}\left(x\right),
\end{split}
\tag{S15}
\end{equation*}
with the $n$-th band Wannier function on the $j$-th site $\phi_{nj}\left(x\right)$. The periodicity of the potentials leads to $t_{s}^{1,1}(j,j\pm1)=t_{s}$ and $t_{so}^{1,1}(j,j\pm1)=\pm(-1)^{j}t_{so}$. Similar to Ref. \cite{Liu2014}, using the local unitary transformation $\hat{\psi}_{j\downarrow}\rightarrow\left(-1\right)^{j}\hat{\psi}_{j\downarrow}$ and performing the Fourier transform, we can write the effective Hamiltonian in momentum space,
\begin{equation*}\label{equation S16}
\hat{H}_{TI}=\sum_{k\in BZ}\left(\begin{array}{cc}
\hat{\psi}_{k\uparrow}^{\dagger} & \hat{\psi}_{k\downarrow}^{\dagger}\end{array}\right)\left[h_{y}\left(k\right)\sigma_{y}+h_{z}\left(k\right)\sigma_{z}\right]\left(\begin{array}{c}
\hat{\psi}_{k\uparrow}\\
\hat{\psi}_{k\downarrow}
\end{array}\right),
\tag{S16}
\end{equation*}
where the summation is over the first Brilloun zone, $h_{y}\left(k\right)=2t_{so}\sin\left(ka\right)$ and $h_{z}\left(k\right)=m_{z}-2t_{s}\cos\left(ka\right)$, with the lattice constant $a=\pi/k_0$. The topological nature of the ground state can be captured by the so-called winding number, which characterizes the rotation of the spin components of the Hamiltonian in the first Brillouin zone. Mathematically, the winding number is defined as~\cite{Li2013}
\begin{equation*}\label{equation S17}
\mathcal{W}=\sum_{\nu,\nu^{'}=y,z}\oint\frac{dk}{4\pi}\epsilon_{\nu,\nu^{'}}\hat{h}_{\nu}^{-1}\left(k\right)\partial_{k}\hat{h}_{\nu^{'}}\left(k\right),
\tag{S17}
\end{equation*}
where $\epsilon_{y,z}=-\epsilon_{z,y}=1$. The winding number is $1$ for a topologically nontrivial state with $m_z<2t_s$, and $0$ for a topologically trivial state with $m_z>2t_s$. It is worth noting that a homeomorphic transformation, like the local unitary transformation above, cannot change the topology of a system. Therefore this winding number also characterizes the topology of the Hamiltonian in Eq.~(\ref{equation S14}).

\subsubsection{\textbf{2. Beyond the single-band tight-binding limit}}
When $\eta_A$ increases further, the cavity-induced hopping term becomes more important, and the single-band tight-binding approximation may not apply. If fact, we will show here that for typical parameters in the superradiant region on the phase diagram (Fig.~3 in the main text), the single-band approximation fails, as the onsite interband hopping becomes dominant.

As shown in Fig.~\ref{fig:hopping}a, the effective lattice depth $V=V_{0}+\xi_{A}\alpha^{\ast}\alpha$ and the effective Raman-field strength $M=\eta_{A}\left(\alpha^{\ast}+\alpha\right)$ are plotted as functions of $\eta_A$ for typical parameters at $m_z=0$. The lowest-band Wannier functions of the lattice Hamiltonian $p_{x}^{2}/2m+V\cos^{2}\left(k_{0}x\right)$ at three adjacent sites are shown in Fig.~\ref{fig:hopping}e as  $\phi_{11}$, $\phi_{12}$ and $\phi_{13}$, respectively. The lattice hopping rates $t_{s}^{n,n^{'}}(i,j)$ (see Eq.~\ref{equation S15}) between different sites are plotted as functions of $\eta_A$ in Fig.~\ref{fig:hopping}b. The Raman-induced hopping rates $t_{so}^{n,n^{'}}(i,j)$ (see Eq.~\ref{equation S15}) are plotted in Fig.~\ref{fig:hopping}c,d. From Fig.~\ref{fig:hopping}b,c, we see that the nearest-neighbor hopping is always much larger than the next-nearest-neighbor hopping.

More importantly, we see in Fig.~\ref{fig:hopping}d that the Raman-induced on-site interband hopping between the lowest band and the first excited band quickly becomes dominant over other hopping processes when $\eta_A$ increases. Hence, a single-band tight-binding approximation cannot be applied to describe the system when $\eta_A$ is large. This conclusion also applies for the cases $m_z\neq0$. Therefore, in the main text, we have numerically diagonalized the effective Hamiltonian (Eq.~(1) in the main text) to include processes beyond the single-band tight-binging approximation.

\begin{figure}
\includegraphics[width=17cm]{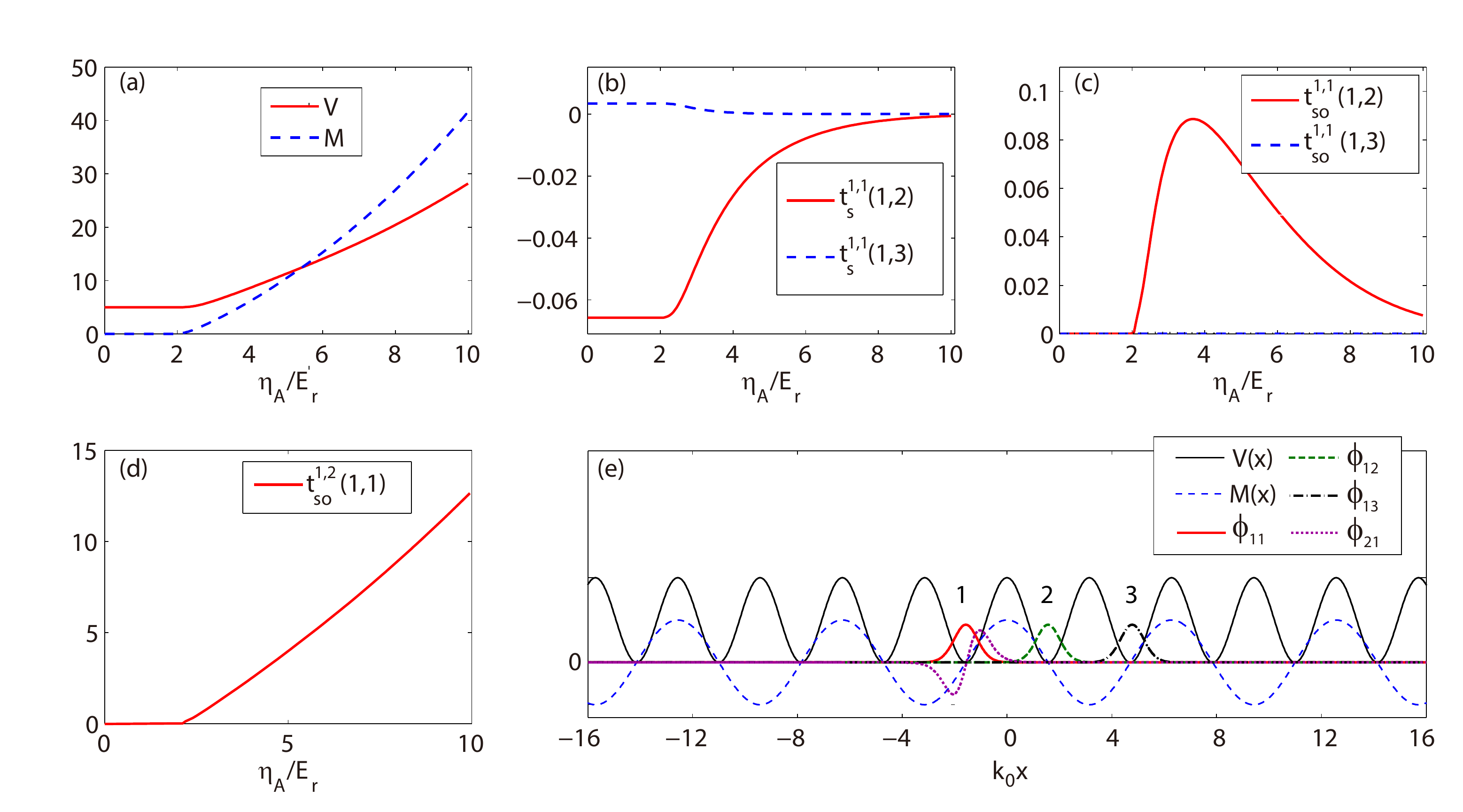}
\caption{(Color online) (a) The effective lattice depth $V=V_{0}+\xi_{A}\alpha^{\ast}\alpha$ and the effective Raman term $M=\eta_{A}\left(\alpha^{\ast}+\alpha\right)$ as functions of $\eta_A$ . (b) The spin-conserving lattice hopping rates between the three adjacent sites (shown in (e)) as functions of $\eta_A$. (c) The spin-flipping Raman-induced hopping rates between adjacent sites as a function of $\eta_A$. (d) The spin-flipping on-site hopping between the lowest band and the first exited band. (e) The effective lattice potential, the effective Raman field, the lowest-band Wannier functions of different sites ($\phi_{11}$, $\phi_{12}$, $\phi_{13}$), and of the first excited band ($\phi_{21}$). The parameters are the same as those in Fig.~2a in the main text. }
\label{fig:hopping}
\end{figure}

\subsubsection{\textbf{3. The calculation of winding number by exactly diagonalizing the effective Hamiltonian }}
Since the single-band tight-binding model is insufficient, we need to numerically evaluate the winding number by exactly diagonalizing the effective Hamiltonian. For that purpose, we introduce the local gauge transformation $\hat{\psi}_{\downarrow}\rightarrow-ie^{ik_{0}x}\hat{\psi}_{\downarrow}$, to map the Hamiltonian of Eq.~(\ref{equation S6}) into its homeomorphic form
\begin{equation*}\label{equation S18}
\hat{H}=\int dx\left(\begin{array}{cc}
\hat{\psi}_{\uparrow}^{\dagger} & \hat{\psi}_{\uparrow}^{\dagger}\end{array}\right)\left(\begin{array}{cc}
\frac{p_{x}^{2}}{2m}+V\left(x\right)+m_{z} & -iM\left(x\right)e^{ik_{0}x}\\
iM\left(x\right)e^{-ik_{0}x} & \frac{\left(p_{x}+k_{0}\right)^{2}}{2m}+V\left(x\right)-m_{z}
\end{array}\right)\left(\begin{array}{c}
\hat{\psi}_{\uparrow}\\
\hat{\psi}_{\downarrow}
\end{array}\right),
\tag{S18}
\end{equation*}
where $V(x)=\left(V_{0}+\xi_{A}\alpha^{\ast}\alpha\right)\cos^{2}\left(k_{0}x\right)$ and $M\left(x\right)=\eta_{A}\left(\alpha^{\ast}+\alpha\right)\cos\left(k_{0}x\right)$. The winding number is read out from the spin texture of the Bloch states at the occupied lowest band, as shown in Fig.~3 in the main text. We map out the topological phase boundaries on the steady-state phase diagram (see Fig.~4 in the main text) by calculating the winding number of the steady state of Eq.~(\ref{equation S18}). The topological phase transition in the phase diagram is also confirmed by the edge-state calculations.

For the single-band tight-binding model, the topology of the system becomes trivial as $m_z$ is larger than $2t_s$~\cite{Liu2014}. Note that in our system, $m_c$ denotes the band width of the background lattice. Since $t_s$ is inversely proportional to the effective lattice depth $V=V_{0}+\xi_{A}\alpha^{\ast}\alpha$, as $\eta_A$ increases, the bandwidth $2t_s$ decreases. Thus, when $m_z<m_c$, the single-band tight-binding model predicts that the system should undergo a topological phase transition when $\eta_A$ increases. When $m_z>m_c$, according to the single-band tight-binding model, there should be no topological phase transitions when $\eta_A$ increases. However, from the steady-state phase diagram generated by fully diagonalizing the Hamiltonian, we see that the topological phase boundary between the trivial SR state and the TSR state is at $m_z>m_c$. We will show in the following subsection that an understanding of the TSR-SR phase transition here can only be achieved by considering the interband hopping terms.

\subsubsection{\textbf{4. The driving force for the TSR-SR phase transition}}

\begin{figure}
\includegraphics[width=17cm]{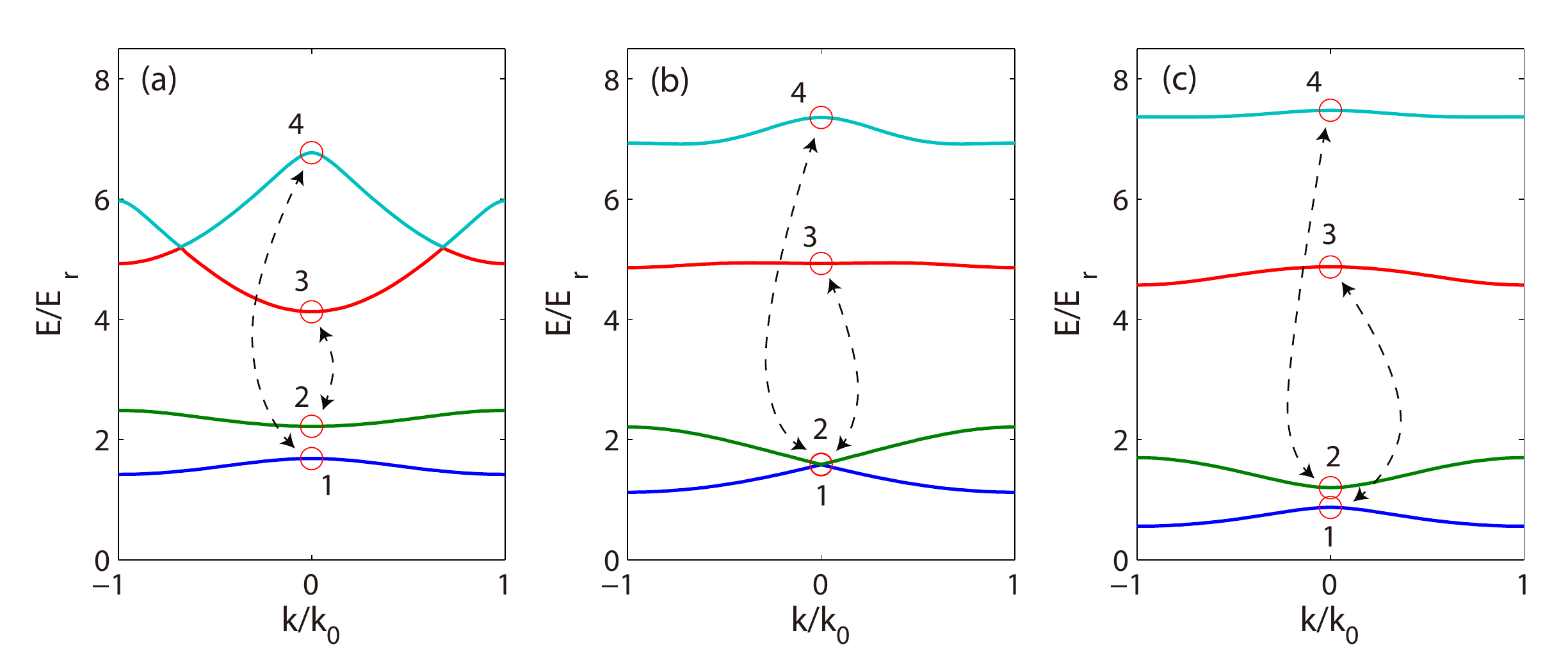}
\caption{(Color online) Band structure and Raman-induced interband couplings, with $m_z=0.4E_r>m_c$, and (a) $\eta_A\sim 3.27E_r$, (b) $\eta_A\sim 3.80E_r$, (c) $\eta_A\sim 4.33E_r$. Other parameters are the same as in Fig.~4 in the main text. The interband couplings among states $1$, $2$,$3$, and $4$ at $k=0$ (red circles) are indicated by arrows. }
\label{fig:band}
\end{figure}

In this subsection, we show that the TSR-SR phase transition is driven by the Raman-induced on-site interband coupling. By diagonalizing the Hamiltonian Eq.~(\ref{equation S18}), we calculate the evolution of the steady-state band structure with increasing $\eta_A$ at a fixed $m_z>m_c$. The TSR-SR transition is characterized by the bulk gap closing and re-opening between the lowest two bands at $k=0$ as $\eta_A$ increases. As shown in Fig.~\ref{fig:band}a, in the I phase, an insulating gap is already opened by the Zeeman field $m_z$ for the Fermi gas at half filling. As $\eta_A$ increases, the system becomes superradiant, and the cavity-assisted Raman field will couple different bands, leading to the deformation of the band structure. Since the Raman field only couples different spin-components, at $k=0$, we have numerically checked that state $1$ is coupled to state $4$, and state $2$ is coupled to state $3$ (see the illustrations in Fig.~\ref{fig:band}a). These interband couplings will push the bands involved toward opposite directions. As a result, states 1 and 2 will be shifted downward while states 3 and 4 will move upward. Meanwhile, since the smaller gap between states 2 and 3 leads to a stronger interband coupling and consequently a wider energy splitting, the downward shift of state $2$ is faster than that of state $1$, as $\eta_A$ and the interband coupling strength increase. These give rise to the bulk gap closing at the TSR-SR phase boundary (see Fig.~\ref{fig:band}b). After the band inversion at the topological phase boundary, state $1$ (state $2$) is coupled to state $3$ (state $4$) (see the illustrations in Fig.~\ref{fig:band}c). Therefore, the bulk gap in the TSR state will increase with increasing $\eta_A$ (and the interband coupling strength). Hence, the TSR-SR phase transition here is driven by the Raman-induced interband coupling.

Interestingly, for $m_z>m_c$, right at the gap closing point (Fig.~\ref{fig:band}b), the Fermi surface at half-filling is reduced to a point. Hence, a spin-up fermion with $k=0$ can be scattered by a cavity photon into a spin-down state at $|k|=k_0$ with the same energy. This nesting condition at the gap closing point leads to a peak in the derivative of the cavity photon number at the topological phase boundary, which can be used as a signal for the detection of the bulk gap closing and the corresponding topological phase boundary. We note that for $m_z>m_c$, the nesting condition is satisfied only at this TSR-SR topological phase boundary; while for $m_z<m_c$, the nesting condition is satisfied along the SR phase boundary.

\subsection{\textbf{D. Robustness analysis of detection schemes}}\label{The experimental consideration}

\begin{figure}
\includegraphics[width=15cm]{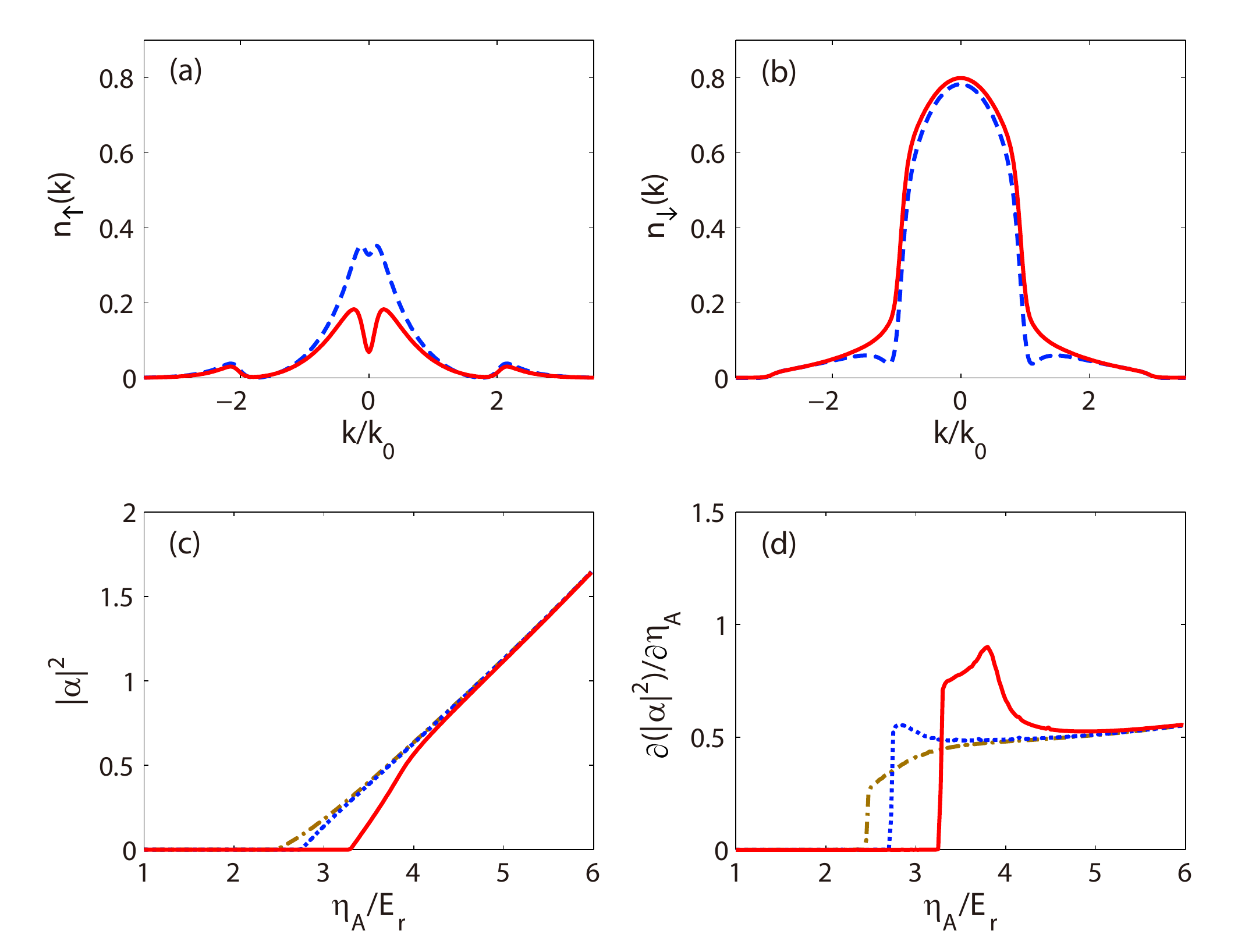}
\caption{(Color online) The detection signals at a realistic temperature and away from half filling, with $k_BT=E_r/30$ and with $76$ atoms on $80$ lattice sites. (a)(b) Momentum distribution of the spin-up (a) and the spin-down (b) atoms in the TSR state (dash) and the trivial SR state (solid), respectively. The Zeeman field for the solid (dash) curves is $m_z=0.255E_r$ ($m_z=0.305E_r$). Here, $\eta_A= 3.5E_r$, and the critical field for the topological phase transition is at $m_z\sim 0.28 E_r$. Other parameters are the same as those in Fig.~5 in the main text.  For $^6$Li atoms, the temperature corresponds to $\sim118$nK. (c)(d) The cavity photon number (c) and its derivative with respect to $\eta_A$ (d) across the topological phase boundary as $\eta_A$ is tuned. The parameters are: $m_z=0.03E_r$ (dash-dotted), $m_z=0.132E_r$ (dash), $m_z=0.4E_r$ (solid). }
\label{fig:signal}
\end{figure}

In the main text, we propose to detect the topological phase transition between the SR phase and TSR phase by measuring the abrupt changes in the atomic momentum distribution or the variation of the cavity photon number as the system crosses the critical point. Now we will analyze the robustness of these signals at higher temperatures and away from half filling.

As an example, we adopt the typical parameters of $^{6}$Li atoms, where the states $|\uparrow\rangle$ and $\downarrow\rangle$ respectively correspond to the states $|F=\frac{1}{2},m_F=\frac{1}{2}\rangle$ and $|F=\frac{1}{2},m_F=-\frac{1}{2}\rangle$ in the ground hyperfine manifold. The recoil energy $E_r$ is then estimated to be $\sim 73.7$kHz with $k_0=2\pi/\lambda$, where $\lambda\sim670$nm such that the transverse pumping laser is blue detuned with respect to the atomic transition. Then, the parameters in Fig.~2 of the main text can be satisfied by choosing: $ T\sim 17.7$nK, $g_A\sim27.1$MHz, $\Delta=2$GHz, and $\kappa\sim7.4$MHz. Under these parameters, we have shown in the main text that the topological phase transition can be detected by measuring the spin-selective momentum distribution or by measuring the cavity photons leaking out of the cavity.

At a higher temperature of $118$nK  (i.e., $k_BT=E_r/30$), and with the Fermi energy slightly deviated from the half-filling condition, we find that both signals are still present (see Fig.~\ref{fig:signal}). One may still detect the topological phase transition by looking at the abrupt change of the momentum space density distribution of spin-down atoms at $k=\pm k_0$ and of spin-up atoms at $k=0$; or by looking at the peak structure in the derivative of the cavity photon number.

We have also checked that with more atoms in the cavity, the requirements on the cavity parameters can be considerably relaxed. For example, if we consider the case where there are $1.6\times 10^4$ atoms in the optical cavity, then at a temperature of $k_BT=E_r/30$ ($T\sim 118$nK for $^6$Li atoms), we can obtain a similar phase diagram, with the tetracritical point located at $\eta_A\sim 0.42E_r$, $m_c\sim 0.132E_r$. For $^6$Li atoms, this may be satisfied by choosing more favorable parameters, e.g., a weaker cavity-atom coupling $g_A\sim 4.3$MHz, a larger single-photon detuning $\Delta=10$GHz, and a larger cavity decay rate $\kappa\sim 22.2$MHz. We emphasize that these parameters can be optimized further.

\end{widetext}


\begin{thebibliography}{99}
\bibitem{footnote1} Under the large-detuning condition, we have assumed $\Delta$ to be the same for both hyperfine states.
\bibitem{footnote2} In the absence of the background lattice ($V_0=0$), there will be only M and TSR states on the phase diagram, due to the absence of the I state.
\end{thebibliography}
\end{document}